\documentclass[12pt,a4paper]{article}
\usepackage{graphicx,hyperref}
\newcommand{\newc}{\newcommand}
\def\u#1{\verb!#1!\endgroup}

\newc{\HW}{\textsf{HERWIG}}
\newc{\TAUOLA}{\textsf{TAUOLA}}
\newc{\ThePEG}{\textsf{ThePEG}}
\newc{\HWPP}{\textsf{Herwig++}}
\newc{\evt}{\textsf{EvtGen}}
\newc{\fortran}{\textsf{FORTRAN}}
\newc{\decayer}{\textsf{Decayer}}

\newc{\HWPPClass}[1]{\href{http://projects.hepforge.org/herwig/doxygen/classHerwig_1_1#1.html}{\textsf{#1}}}
\newc{\ThePEGClass}[1]{\href{http://projects.hepforge.org/thepeg/doxygen/classThePEG_1_1#1.html}{\textsf{#1}}}
\newc{\HWPPParameter}[2]{\href{http://projects.hepforge.org/herwig/doxygen/#1Interfaces.html\##2}{{\bf #2}}}
\newc{\ThePEGParameter}[2]{\href{http://projects.hepforge.org/thepeg/doxygen/#1Interfaces.html\##2}{{\bf #2}}}
\newc{\HWPPParameterValue}[3]{\href{http://projects.hepforge.org/herwig/doxygen/#1Interfaces.html\##2}{{\bf [#2=#3]}}}
\newc{\ThePEGParameterValue}[3]{\href{http://projects.hepforge.org/thepeg/doxygen/#1Interfaces.html\##2}{{\bf [#2=#3]}}}

\begin{document}
\tolerance=100000
\thispagestyle{empty}
\setcounter{page}{0}
 \begin{flushright}
Cavendish-HEP-07/10\\
CERN-PH-TH/226\\
CP3-07-30\\
IPPP/07/89\\
DCPT/07/178\\
KA-TP-30\\
November 2007
\end{flushright}
\begin{center}
{\Large \bf Herwig++ 2.1 Release Note}\\[0.7cm]

{M.~B\"ahr$^1$\\[0.4mm]
E-mail: \tt{mb@particle.uni-karlsruhe.de}}\\[4mm]
{S.~Gieseke$^{1,2}$\\[0.4mm]
E-mail: \tt{gieseke@particle.uni-karlsruhe.de}}\\[4mm]
{M.~Gigg$^3$\\[0.4mm]
E-mail: \tt{m.a.gigg@durham.ac.uk}}\\[4mm]
{D.~Grellscheid$^3$\\[0.4mm]
E-mail: \tt{David.Grellscheid@durham.ac.uk}}\\[4mm]
{K.~Hamilton$^4$\\[0.4mm]
E-mail: \tt{hamilton@fyma.ucl.ac.be}}\\[4mm]
{O.\ Latunde-Dada$^5$\\[0.4mm]
E-mail: \tt{seyi@hep.phy.cam.ac.uk}}\\[4mm]
{S.\ Pl\"atzer$^1$\\[0.4mm]
E-mail: \tt{sp@particle.uni-karslruhe.de}}\\[4mm]
{P.\ Richardson$^{2,3}$\\[0.4mm]
E-mail: \tt{Peter.Richardson@durham.ac.uk}}\\[4mm]
{M.~H.\ Seymour$^{2,6}$\\[0.4mm]
E-mail: \tt{Michael.Seymour@cern.ch}}\\[4mm]
{A.\ Sherstnev$^{5}$\\[0.4mm]
E-mail: \tt{cherstn@hep.phy.cam.ac.uk}}\\[4mm]
{B.~R.\ Webber$^{5}$\\[0.4mm]
E-mail: \tt{webber@hep.phy.cam.ac.uk}}\\[4mm]

$^1$\it Institut f\"ur Theoretische Physik, Universit\"at Karlsruhe.\\[0.4mm]
$^2$\it Physics Department, CERN.\\[0.4mm]
$^3$\it IPPP, Department of Physics, Durham University. \\[0.4mm]
$^4$\it Centre for Particle Physics and Phenomenology, Universit\'e Catholique de Louvain.\\[0.4mm] 
$^5$\it Cavendish Laboratory, University of Cambridge.\\[0.4mm]
$^6$\it School of Physics and Astronomy, University of Manchester\\[0.4mm]
\end{center}

\vspace*{\fill}

\begin{abstract}{\small\noindent
    A new release of the Monte Carlo program \HWPP\ (version 2.1) is now
    available. This version includes a number of significant improvements including:
    an eikonal multiple parton-parton scattering model of the underlying event;
    the inclusion of Beyond the Standard Model~(BSM) physics; and a new hadronic
    decay model tuned to LEP data. This version of the program is now fully ready for
    the simulation of events in hadron-hadron collisions.
}
\end{abstract}

\tableofcontents
\setcounter{page}{1}

\section{Introduction}

The last major public version (2.0) of \HWPP, described briefly
in~\cite{Gieseke:2006ga}, was based on the original version~(1.0)
which was reported on in detail in~\cite{Gieseke:2003hm}. There have 
however been significant changes since the original release~(1.0) and
therefore we intend to publish a full manual in the near future. This
release note therefore only discusses the changes which have been made
since the last release~(2.0).

Please refer to \cite{Gieseke:2003hm} and the present paper if
using version 2.1 of the program and the full manual once it is available.

The main new features of this version are an eikonal multiple parton-parton
scattering model of the underlying event based on~\cite{Butterworth:1996zw};
the simulation of BSM physics including the CP-conserving Minimal Supersymmetric Standard
Model~(MSSM), the minimal Universal Extra Dimensions~(UED) model and the Randall-Sundrum
model; a new model of meson and tau decays which has been tuned to 
LEP and B-factory data. In addition a number of other changes, such as the inclusion of 
intrinsic transverse momentum in hadron-hadron collisions, have been
made and a number of bugs have been fixed.

\subsection{Availability}
The new program, together  with other useful files and information,
can be obtained from the following web site:
\begin{quote}\tt
       \href{http://projects.hepforge.org/herwig/}{http://projects.hepforge.org/herwig/}
\end{quote}
  In order to improve our response to user queries, all problems and requests for
  user support should be reported via the bug tracker on our wiki. Requests for an
  account to submit tickets and modify the wiki should be sent to 
  {\tt herwig@projects.hepforge.org}.

  \HWPP\ is released under the GNU General Public License (GPL) version 2 and 
  the MCnet guidelines for the distribution and usage of event generator software
  in an academic setting, which are distributed together with the source, and can also
  be obtained from
\begin{quote}\tt
 \href{http://www.montecarlonet.org/index.php?p=Publications/Guidelines}{http://www.montecarlonet.org/index.php?p=Publications/Guidelines}
\end{quote}

\section{Multiple Interactions}

A dynamic model of the underlying event, in the form of an eikonal multiple
parton-parton scattering model, is now included in \HWPP. It is intended to
provide the same functionality as \fortran\ \HW~\cite{Corcella:2000bw,Corcella:2002jc}
running with \textsf{JIMMY}~\cite{Butterworth:1996zw}.
It is based on the same model \cite{Butterworth:1996zw} but
implements the physics in a way that is somewhat closer to the eikonal
approach than the original version.

\begin{figure}[!hp]
{
  \begin{center}
    \includegraphics[scale=0.7]{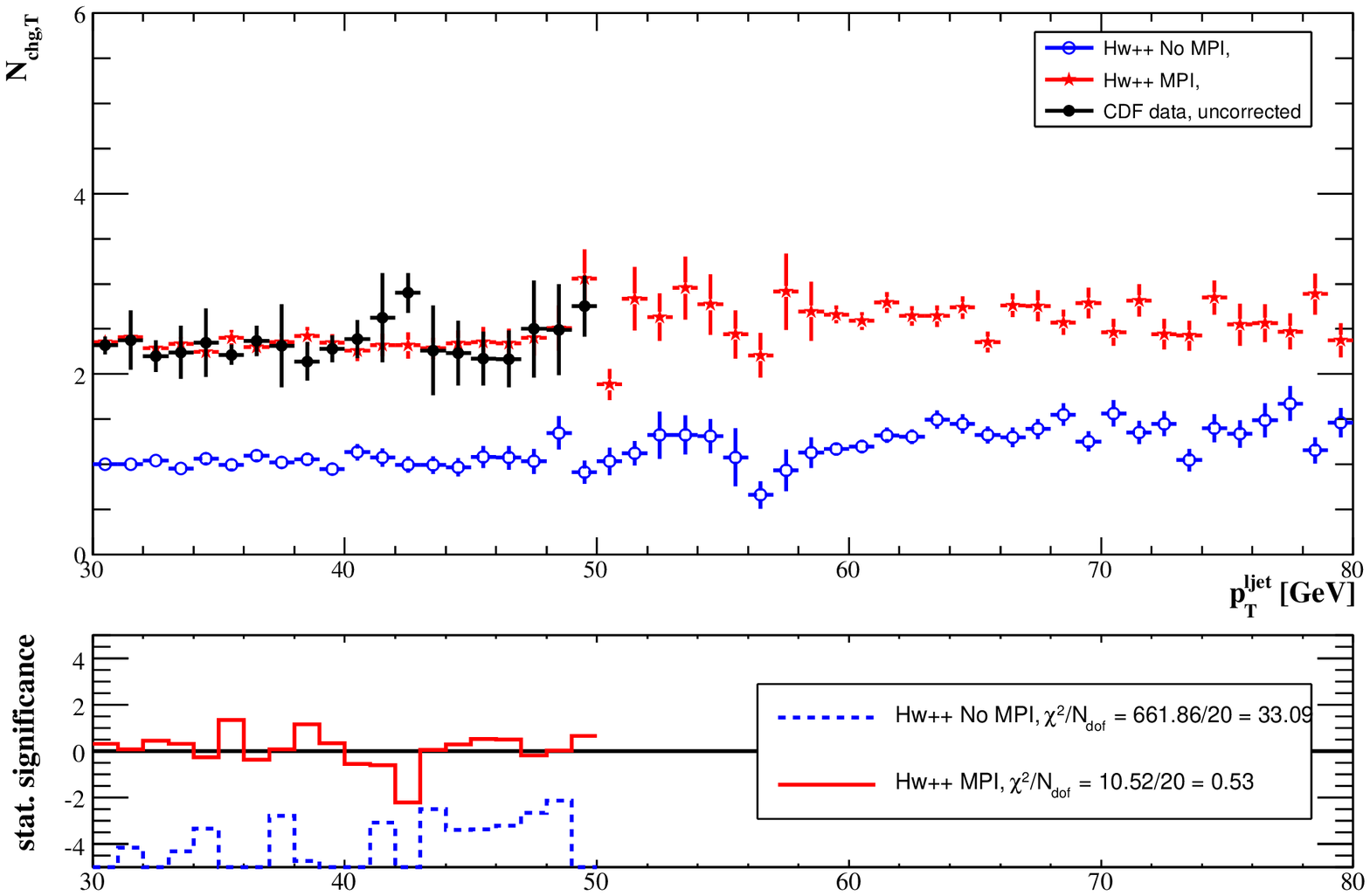}
    \\[1cm]
    \includegraphics[scale=0.7]{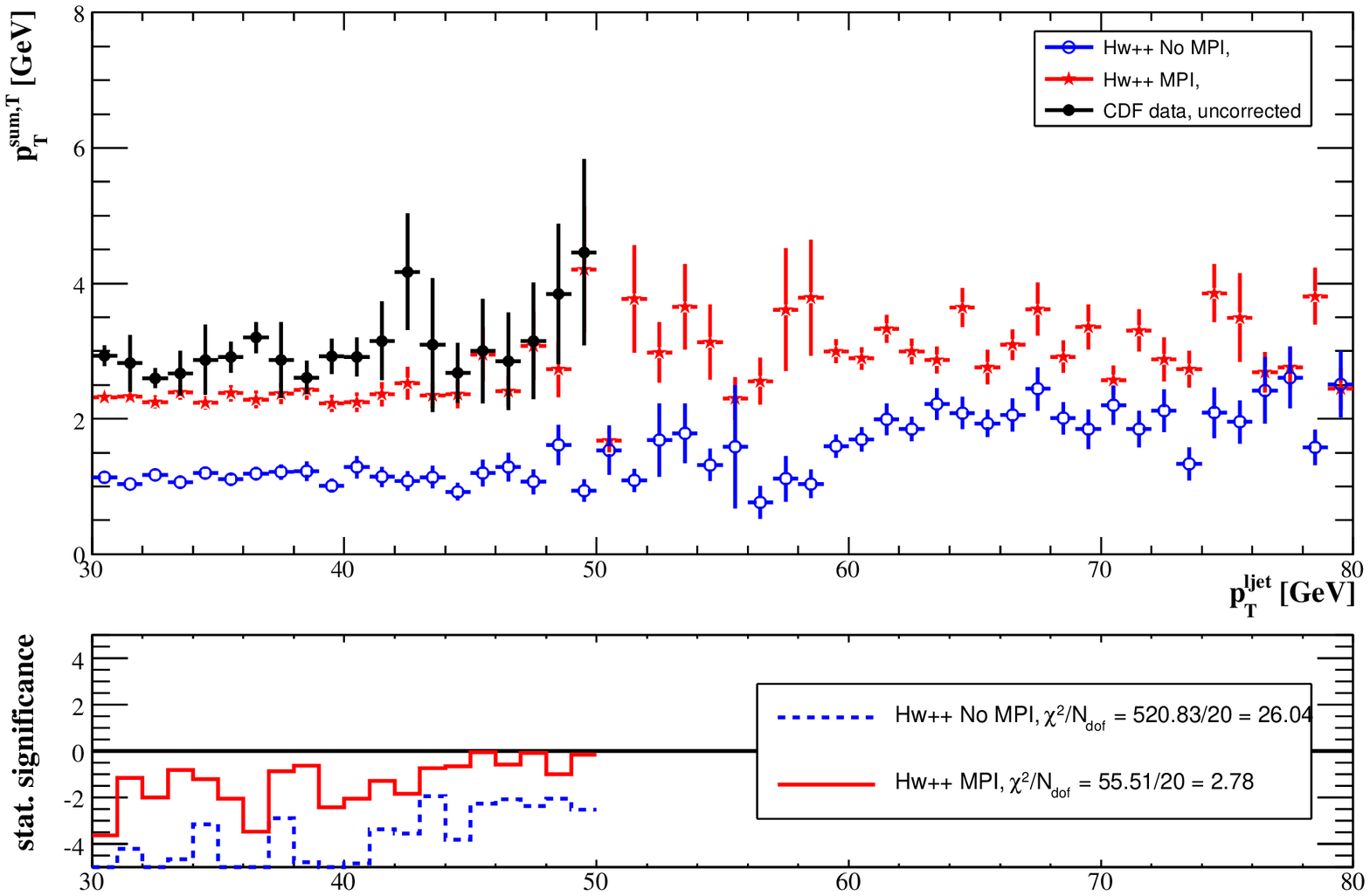}
  \end{center}}
\caption{Average number and average scalar $p_T$ sum of charged particles in
  the transverse region, as defined in \cite{Affolder:2001xt}. The lower parts
  of the plots show the statistical significance of the disagreement between
  prediction and data. Black (full circles) indicates the data points, blue
  (open circles) shows \HWPP~ without multiple parton interactions (MPI) and
  red (stars) shows \HWPP~ with MPI.}
\label{fig:transv}
\end{figure}

The multiple interactions are generated as part of the parton-shower phase of
the event.
After the hard process has been produced and showered the number of
additional scatters in this particular event is calculated. The calculated number of
additional subprocesses are then created and successively processed by the parton
shower. The initial partons from all subprocesses are extracted from the
incoming beam particles, as long as there is enough energy available in the
beams. 

This means, of course, that several colour lines end on the beam particles 
and, furthermore, several quarks are possibly extracted from the beam
particles. This and the prerequisite that the cluster hadronization model expects
only (anti)quarks and (anti)diquarks to hadronize, makes it necessary to force
the initial-state shower to end on a valence quark for the hard subprocess and
to end on a gluon for the additional interactions. After extraction of all additional 
scatters the colour connections between them are set so that the initial
partons of each scattering are connected to the next one. The last pair
are connected to the hadron remnants, which are diquarks if the beam particles
are baryons.

Predictions from the new model as well as \HWPP\ without multiple
parton-parton scattering are shown in Fig \ref{fig:transv}. The predictions of
the Monte Carlo simulations are compared to CDF data~\cite{Affolder:2001xt} 
for two observables, which are sensitive to underlying event activity.  

A more detailed explanation of the model and the data comparison is available
in \cite{MPI}.

\section{BSM Physics}

  The simulation of physics Beyond the Standard Model~(BSM) is now included in
  \HWPP. Rather than implement the matrix elements and decays for a specific
  model, as was done in the \fortran\ program, we have adopted a more general 
  approach~\cite{Gigg:2007cr} which makes extensive use of the \textsf{Helicity}
  libraries in \ThePEG.

  The new framework has the matrix elements for the $2\to2$ hard scattering processes
  and $1\to2$ decays implemented for the different possible spins of the 
  interacting particles. The diagrams are built from \textsf{Vertex} classes,
  each of which is based upon a certain Lorentz structure, which have the
  ability to calculate
  matrix elements and off-shell \textsf{WaveFunctions} for that
  combination of spins. The Feynman rules for a specific model are then 
  implemented by inheriting from these general classes.

  The \textsf{Vertices} for the Minimal Supersymmetric Standard Model~(MSSM),
  minimal Universal Extra Dimensions~(UED) and Randall-Sundrum models are 
  currently implemented. An interface to read files in the SUSY Les Houches
  Accord format~\cite{Skands:2003cj} is included for the MSSM. Once the
  second version of the accord is finalised~\cite{Allanach:2005kk}
  we intend to include additional
  models, such as the Next to Minimal Supersymmetric Standard Model~(NMSSM).

\section{Hadronic Decays}

  A number of major improvements to the simulation of meson and tau decays have been
  made~\cite{Grellscheid:2007tt,MesonDecays}.
  Previously, in \HWPP\ 2.0, the particle properties and decay modes
  for all the particles were the same
  as those used in \fortran\ \HW. Furthermore, 
  for the majority of the hadron decays,  the momenta of the decay products
  were assumed to be uniformly distributed in phase space, or, in 
  a limited number of cases, according to simple weak $V-A$ matrix elements.

  We have updated the properties of the leptons, quarks and mesons to 
  agree with those in~\cite{Yao:2006px} together with some additional
  interpretation and assumptions where necessary. The particle properties
  now used in \HWPP\ can be obtained from
\begin{quote}\tt
       \href{http://www.ippp.dur.ac.uk/~richardn/particles}{http://www.ippp.dur.ac.uk/$\sim$richardn/particles}
\end{quote}
  More importantly, we now use matrix elements and include spin correlations
  for the decays including, where possible, a sophisticated treatment of off-shell
  effects. The new hadron decay model is described in more detail 
  in~\cite{Grellscheid:2007tt,MesonDecays}.

  While the properties of the baryons are currently identical to those used in
  the previous version we have made some changes to ensure that excited $\Xi_b$
  baryons can decay. The masses and decay modes of the $\Xi_b$, $\Xi'_b$ and $\Xi^*_b$
  baryons have been adjusted so that there is only a radiative decay mode
  for the $\Xi'_b$ and a pionic mode for the $\Xi^*_b$.
  The properties of the remaining baryons will be updated in the near future.

\section{Other Changes}

A number of other more minor changes have been made.
The following changes have been made to improve the physics 
simulation:
\begin{itemize}
\item The option of intrinsic transverse momentum has been added. Since this cannot
      be calculated perturbatively, we have to model this using a non-perturbative
      distribution. The distributions that can be used in \HWPP\ are a Gaussian,
      an inverse quadratic or a combination of both. The best fit obtained to Drell Yan $Z$ and $W$
      boson production data at the Tevatron  is a Gaussian
      distribution with a root-mean-square transverse momentum of $2.2$ GeV. Assuming a
      logarithmic dependence on the beam energy, the corresponding value estimated for the
      LHC is $5.7$ GeV.  
\item A number of features for CKKW~\cite{Catani:2001cc} matching have been
      added. The structure is designed to be very flexible, allowing the
      straightforward
      implementation of different matching approaches following the general idea
      of the original proposal. Implementing a new merging prescription amounts 
      to supplying
      the key ingredients of a CKKW-type algorithm, {\it i.e.} a jet measure to reconstruct
      a parton shower history and a jet resolution to separate regions of jet production
      and jet evolution. No approximation is performed in the Sudakov form factors
      used for reweighting.
      Testing and development of an adaptation of the CKKW method
      to the improved angular ordered shower in \HWPP\ is currently underway.

\item A major clean-up and restructuring of the hadronization module has been performed.
      As part of this process a 
      number of hadronization parameters which either did not depend
      on the flavour of the partons forming a cluster, or only had different values for clusters
      which contained a bottom quark, are now different for clusters containing
      light, charm and bottom quarks. These include 
      \HWPPParameter{ClusterFissioner}{ClMax},
      \HWPPParameter{ClusterFissioner}{ClPow},
      \HWPPParameter{ClusterFissioner}{PSplit},
      \HWPPParameter{ClusterDecayer}{ClDir} and
      \HWPPParameter{ClusterDecayer}{ClSmr}. In addition, the option for increasing the 
      threshold for the single hadron decay of bottom clusters above the
      threshold for the production of two hadrons has been extended to 
      charm clusters.

\item  The choice of the $n$ reference vector for initial-state radiation where
       the colour partner of the incoming particle is in the final state has been
       changed. Previously a vector backwards to the direction of the radiating
       particle was chosen in the laboratory frame, whereas we now use the Breit
       frame~\cite{Gieseke:2003rz}.

\item  \ThePEG\ has moved to a templated solution to ensure that the dimensions
       of all calculations are correct, by default this is switched
       off for faster compilation. As part of this change \ThePEG\ now uses
       an internal library for vectors and Lorentz transformations. This means
       that \HWPP\ no longer explicitly depends on \textsf{CLHEP}. \textsf{CLHEP}
       is now only needed when interfacing to external packages, such as \textsf{KtJet},
       which use \textsf{CLHEP}.

\item  The majority of the \textsf{Helicity} classes which were previously in \HWPP\
       have been moved to the \ThePEG. The classes for the calculation of the 
       \textsf{WaveFunctions} and the general \textsf{Vertices} based on
       spin structures have been moved to \ThePEG. The classes which implement
       the vertices for specific models remain in \HWPP\ but 
       have been moved to the relevant model directory. The \textsf{Helicity} directory
       no longer exists.

\item The shower and hadronization modules have been extended to handle the 
      showering and hadronization of processes which violate baryon number
      conservation.

\item  Additional options for the non-perturbative behaviour of the strong coupling
       constant used in the shower have been added to the \HWPPClass{ShowerAlphaQCD}
       class.

\item  The structure of the input files has been significantly modified. In addition
       a default version of the \ThePEGClass{Repository} file is now installed
       so that most users will only need to use the \textsf{read} and \textsf{run}
       stages of the \HWPP\ program. A number of changes to \textsf{Switches} have also 
       been made to improve the consistency of the names and options.
       The name of all the classes in the \ThePEGClass{Repository} has been 
       changed from \textsf{Herwig++::} to \textsf{Herwig::} for consistency as well.

\item  Changes have been made to ensure that no radiation harder than the 
       scale of the hard process
       is emitted in the parton shower, as was the case in \fortran\ \HW.

\item The non-perturbative gluon splitting in the \HWPPClass{PartonSplitter} class
      now decays the gluons to all the
      quark-antiquark pairs which are kinematically allowed, with the probability
      of producing a given quark-antiquark pair proportional to the available
      phase space. Previously $u\bar{u}$ and $d\bar{d}$ pairs were produced with
      equal probabilities, however there is no change in behaviour for the 
      default values of the parameters.

\item The option of forbidding the production of specific hadrons during the
      hadronization has been added in order to forbid the production of
      $\sigma$ and $\kappa$ mesons, which are included to represent $s$-wave systems
      in some meson decays.

\item  The default parton distribution function~(PDF) has been changed to
       the leading-order set of~\cite{Martin:2002dr}. 
       A change has been made so that, rather than the cubic interpolation previously used
       for the internal MRST PDFs, we now switch to linear interpolation above $x=0.8$
       to give improved behaviour at high-$x$. 

\item The Higgs boson running width is now calculated as described 
      in~\cite{Seymour:1995qg}. The \HWPPClass{SMHiggsMassGenerator} class has been added
      to generate the off-shell mass of the Higgs boson using the prescription of~\cite{Seymour:1995qg}
      and is used to give the mass distribution of the Higgs boson in $gg,q\bar{q}\to h^0$ and Higgs boson plus
      jet processes. The Higgs boson branching ratios are now calculated for each decaying Higgs boson, based
      on its off-shell mass, and therefore the values in the data tables are not used. If modes are
      switched off the \HWPPClass{SMHiggsMassGenerator} class correctly takes this into account
      when used as part of the cross-section calculation.

\item An interface to \textsf{ROOT}~\cite{Brun:1997pa} has been added so that it
      is easier to use  \textsf{ROOT} in \ThePEGClass{AnalysisHandler}s if desired.

\item A number of additional \ThePEGClass{AnalysisHandler}s have been added comparing the
      results of \HWPP\ to \textsf{LEP} and B-factory data. These were used to tune the shower and
      hadronization parameters in the current version of the program.

\item The forced splitting required to ensure the correct valence content of the 
      incoming hadrons, which was previously preformed as the first step of the
      hadronization, is now performed as the last step of the shower. This is part of
      the changes made to include multiple parton-parton scattering.

\item The handling of partonic decays of bottom and charm hadrons has been changed.
      The \textsf{PartonicHadronizer} class, which was previously used to perform
      the hadronization of these decays, has been replaced by functionality in
      the \HWPPClass{PartonicDecayerBase} class from which all classes implementing
      partonic decays now inherit.

\item The shower module has been changed so that in the showering of particles
      which have already been decayed the off-shell mass of the particle is preserved,
      rather than the particle being given its on-shell mass at the end of the shower,
      in order to ensure the conservation of energy and momentum in the parton shower.

\item An interface to the \evt~\cite{Lange:2001uf} decay package has been added
      including spin correlation effects. This interface currently needs a modified
      version of \evt\ which is available on request but we hope these changes will
      be included in the LHC version of \evt\ in the near future.

\item  A new class, \HWPPClass{MEQCD2to2Fast}, for QCD $2\to2$ scattering
       processes which uses hard-coded expressions for the matrix elements
       rather than the helicity libraries of \ThePEG\ has been added for
       use in the multiple parton-parton scattering model of the underlying event to
       increase the speed of the model.

\item The {\tt herwig-config} script has been added to the release to give information
      on the installed location of the program, etc., for use by other programs, such as
      \textsf{RivetGun} which use \HWPP.

\item Options to handle the decay of a colour-singlet particle to three
      quarks or antiquarks, or the decay of a colour triplet particle to 
      two antiquarks, have been added to the \HWPPClass{MamboDecayer} class
      in order to test the showering and hadronization of systems which violate
      baryon number conservation. 

\item  The implementation of the veto used to include the effects of the parton
       distribution functions~(PDFs) in the backward evolution algorithm used to
       generate the initial-state parton shower has been modified to increase the
       efficiency of the veto, and hence the speed of the algorithm.

\item A general interface for vetoing parton shower emissions has been added.
      Vetos on a single emission, a shower attempt or the whole event are
      possible. One implementation of this interface provides vetoing branchings
      depending on the $p_\perp$ of the branching.

\item  A default rapidity cut of $|y|<3$ has been imposed for photons produced
       in the hard process to reduce the number of events at high rapidity
       and small-$x$ which could not be successfully showered.

\item  \HWPP\ now uses the GNU scientific library~(GSL)~\cite{GSL}
       for some special mathematical functions.


\item A number of improvements have been made to the \textsf{DOXYGEN} documentation.

\end{itemize}

The following bugs have been fixed:
\begin{itemize}
\item A bug in identifying which partons in clusters came from the perturbative
      stage of the event has been fixed. This meant that the wrong hadrons preserved the 
      direction of their parton constituents when using the
      \HWPPParameterValue{ClusterDecayer}{ClDir}{1} option.

\item A bug affecting the calculation of the soft cluster masses for clusters containing
      a remnant of the beam particle has been fixed.

\item A number of changes have been made to improve the identification of
      particles which have already been decayed but should be showered, for example in 
      BSM physics processes or for processes supplied using the Les Houches accord.

\item A bug affecting the generation of the phase space in the \HWPPClass{MamboDecayer}
      class has been fixed.

\item Several filenames have been changed to avoid problems with case-insensitive
      Mac OS X file systems.

\item A bug preventing the use of the \ThePEGClass{NoPDF} parton distribution function
      in lepton-lepton collisions has been fixed.
\end{itemize}

\section{Summary}

  \HWPP2.1 is the second version of the \HWPP\ program with a complete simulation of 
  hadron-hadron physics albeit with major improvements to the simulation of 
  the underlying event and Beyond the Standard Model physics with respect to 
  the previous version. The program has been extensively tested against
  a large number of observables from LEP, Tevatron and B factories.
  All the features needed for realistic studies for 
  hadron-hadron collisions are now present and  we look forward to 
  feedback and input from users, especially
  from the Tevatron and LHC experiments.

  Our next major milestone is the release of version 3.0 which will be at least as
  complete as \HW\ in all aspects of LHC and linear collider simulation.
  Following the release of \HWPP3.0 we expect that support for the 
  {\sf FORTRAN} program will cease.

\section*{Acknowledgements} 

  This work was supported by Science and Technology
  Facilities Council, formerly the Particle Physics and Astronomy Research Council,
  and the European Union Marie Curie Research Training Network MCnet under
  contract MRTN-CT-2006-035606. The research of K.~Hamilton was supported by 
  the Belgian Interuniversity Attraction Pole, PAI, P6/11. 
  M.~B\"ahr acknowledges support from the ``Promotionskolleg am Centrum
  f\"ur Elementarteilchenphysik und Astroteilchenphysik CETA'' and
  Landesgraduiertenf\"orderung Baden-W\"urttemberg.
  S.~Pl\"atzer acknowledges support from the Landesgraduiertenf\"orderung
  Baden-W\"urttemberg.
  S.~Gieseke and P.~Richardson 
  would like to thank the CERN Theoretical Physics and Physics Software groups for their
  hospitality. We would like to thank A.~Ribon and P.~Stephens for their
  contributions to the \HWPP\ project and previous versions of the program.

\providecommand{\href}[2]{#2}\begingroup\raggedright\endgroup

\end{document}